\documentclass[10pt,aps,prd,nofootinbib,superscriptaddress, twocolumn,preprintnumbers]{revtex4}
\usepackage{latexsym,amssymb,amsmath,graphics,graphicx, hyperref}

\def\LL{{\cal L}}

\newcommand{\TeV}{\,\mathrm{TeV}}
\newcommand{\GeV}{\,\mathrm{GeV}}

\newcommand{\MET}{\mbox{$E_T\hspace{-0.225in}\not\hspace{0.18in}$}}
\newcommand{\smMET}{\mbox{$E_T\hspace{-0.195in}\not\hspace{0.13in}$}}

\newcommand{\gsim}{\lower.7ex\hbox{$\;\stackrel{\textstyle>}{\sim}\;$}}
\newcommand{\lsim}{\lower.7ex\hbox{$\;\stackrel{\textstyle<}{\sim}\;$}}

\begin{document}

\title{It's On: Early Interpretations of ATLAS Results in Jets and Missing Energy Searches}

\author{Daniele S. M. Alves}
\affiliation{Theory Group, SLAC National Accelerator Laboratory, Menlo Park, CA 94025}
\affiliation{Physics Department, Stanford University, Stanford, CA 94305}

\author{Eder Izaguirre}
\affiliation{Theory Group, SLAC National Accelerator Laboratory, Menlo Park, CA 94025}
\affiliation{Physics Department, Stanford University, Stanford, CA 94305}

\author{Jay G. Wacker}
\affiliation{Theory Group, SLAC National Accelerator Laboratory, Menlo Park, CA 94025}

\begin{abstract}
The first search for supersymmetry from ATLAS  with 70 $\text{nb}^{-1}$ of integrated luminosity extends the Tevatron' s reach for colored particles that decay into jets plus missing transverse energy.
For gluinos that decay directly or through a one step cascade into the LSP and two jets, the mass range $m_{\tilde{g}} \le 205 \GeV$ is disfavored by the ATLAS searches, regardless of the mass of the LSP.  In some cases the coverage extends up to $m_{\tilde{g}} \simeq 295 \GeV$, already surpassing the Tevatron's reach for compressed supersymmetry spectra.
\end{abstract}
\preprint{SLAC-PUB-14219}
\preprint{arXiv:1008.0407}
 \maketitle

\section{Introduction}
Jets plus missing energy is one of the most promising  channels in the search for physics beyond the Standard Model at hadron colliders.  With the preliminary results from ATLAS on jets plus missing energy searches for supersymmetry \cite{ATLASSusy},
new limits on color octet particles that subsequently decay into a pair of jets and a long lived neutral particle can be set.
Although only $70\pm 8\text{ nb}^{-1}$ of integrated luminosity has been analyzed, that is already sufficient to infer new constraints on models with light colored particles.

The most robust bounds on new color octet particles come from LEP2's measurements of thrust.  These measurements place a model independent lower bound on a new color octet particle of 51 GeV \cite{Kaplan:2008pt}.    Above this energy, the limits are model dependent and derive from searches that trigger on jets and missing energy.   In fact, most recent searches have focused on models closely related to mSUGRA \cite{CMSSM}.  Within the many assumptions of mSUGRA lies the condition of gaugino mass unification.  This requirement imposes  that throughout all of the accessible mSUGRA parameter space, the LSP is a bino, the fermionic partner of the hypercharge gauge boson, and the gluino is approximately six times heavier than the bino (see \cite{Tevatron} for recent Tevatron searches).  
Therefore, exclusion limits that are stated in the mSUGRA parameter space never effectively vary the kinematics of the signal.\footnote{UA2 provides some of the most extensive discussion on gluino limits away from gaugino mass unification \cite{UA2}.}

Even within supersymmetric models with gauge coupling unification, gaugino mass unification is not an automatic outcome.
For instance, in mirage mediation which mixes gravity mediation with anomaly mediation, the gauginos can be degenerate at the TeV scale \cite{MirageMediation}.   In non supersymmetric models that also have a gluino-like object in their spectrum, such as Universal Extra Dimension models (UEDs) with the lightest KK gluon, gaugino mass unification is typically not satisfied \cite{UED}.  
Therefore, most jets plus missing energy studies do not directly apply to these theories.

Moreover, in mSUGRA parameterizations the gluino and squarks' masses are related, and analyses based on such parameterizations rely on the production of squarks as well as gluinos to place indirect bounds on the gluino mass.

Model independent studies reinterpreted the Tevatron's searches for jets and missing energy in a more general setting \cite{TevatronJetsMET}.  Models where the gluino decays into jets plus the LSP have a universal lower bound of $m_{\tilde{g}}\gsim130\GeV$ (stronger constraints apply to spectra where the gluino and LSP have hierarchical mass splittings).  This universal limit is far below the normally quoted 400 GeV limit on gluino-like objects.  
Given such modest current bounds -- a 51 GeV hard lower limit on the gluino mass and an approximate 130 GeV lower limit from dedicated searches -- the integrated luminosity accumulated at the LHC in its early running is already sufficient to extend the present reach for light colored octets.   



Low luminosity searches allow modest cuts to be placed on missing transverse energy, helping the searches' efficiencies for discovering compressed spectra.   
ATLAS performed four separate searches for excesses in jets plus missing energy: monojets, dijets plus missing energy, three jets  plus missing energy and four jets plus missing energy.    These searches required a leading jet with $p_T > 70 \GeV$ and $\MET > 40\GeV$.    This low $\MET$ cut is particularly sensitive to theories with somewhat compressed spectra and improves over the Tevatron's limits.

 \section{Model and Simulation}
 
 This article considers a limit of the MSSM where the gluino is the only accessible colored particle, and decays via an off-shell
 squark into two jets plus the lightest supersymmetric particle (LSP).
 More specifically, the theory consists of a color octet Majorana fermion, $\tilde{g}$,  with no $SU(2)_L \times U(1)_Y$ charges, and
a lighter Majorana fermion with no Standard Model gauge quantum numbers, $\chi^0$.
 $\chi^0$ is stable on detector timescales and may be the dark matter of the Universe.   $\tilde{g}$ can decay
 into $\chi^0$ plus two jets or to $\chi^0$ and a gluon.  For simplicity, it will be assumed that the branching ratio into light quarks is  unity, $\text{Br}(\tilde{g}\rightarrow \chi^0 q\bar{q}) =1$.    The decay is mediated by a dimension 6 operator
 \begin{eqnarray}
\LL_{\text{decay}} = \frac{1}{\Lambda^2}  (\bar{\chi^0}q_a)\, T^A{}^a_b\,(\bar{q}^b \tilde{g}^A),
  \end{eqnarray}
which can be generated by integrating out a color triplet scalar.   The lifetime of $\tilde{g}$ is approximately
\begin{eqnarray}
\Gamma_{\tilde{g}} \simeq \frac{ (m_{\tilde{g}} - m_{\chi^0})^5}{ 4\pi \Lambda^4}
\end{eqnarray}
 which leads to a prompt decay so long as  $\Lambda \lsim 10 \TeV$ for $m_{\tilde{g}} \gsim m_{\chi^0} + 10 \GeV$.   There is no {\em a priori} relation between  the masses of $\chi^0$ or $\tilde{g}$.   $\chi^0$ may be very light without any constraints arising  from  LEP, and the only model independent constraint on $\tilde{g}$ is that it should be heavier than 51 GeV.

Models with approximate  $\tilde{g}$-$\chi^0$ mass degeneracy are particularly challenging for jet plus $\MET$ searches. 
In this case,  the decay products of $\tilde{g}$ are soft and buried in background.  
In the degenerate limit, the most efficient way to detect  $\tilde{g}$ production is by looking for radiation of additional jets.    At the Tevatron,  pair produced $\tilde{g}$'s plus radiation gives rise to events with low multiplicity jets plus $\MET$.  In particular, monojet searches can be effective at discovering these topologies \cite{TevatronJetsMET}.  However, monojet searches are typically exclusive and place poor bounds away from the degenerate limit.  For instance, CDF places a second jet veto of $E_{T\, j_2} \le 60 \GeV$ and a third jet veto of $E_{T\, j_3} \le 20 \GeV$  \cite{CDFMonojet}.
As the mass difference between $\tilde{g}$ and $\chi^0$ increases, the efficiency of such cuts diminishes.    In the non-degenerate limit, the most suitable searches have higher jet multiplicity. However, the cuts applied on the monojet and multijet searches performed  so far were sufficiently strong  that  they left a gap in the coverage of the intermediate mass-splitting region \cite{TevatronJetsMET}.  The present bound on $m_{\tilde{g}}$  only extends above 130 GeV for $m_{\chi^0}\lsim100\GeV$.  The LHC cross section for gluinos just above this limit is of the order of a few nanobarns. Therefore, limits can be  improved by the LHC with remarkably low luminosity and early discovery is potentially achievable.   Unfortunately, no excesses were observed in \cite{ATLASSusy}, so only new limits could be inferred.

\begin{figure}[htbp]
\begin{center}
\includegraphics[width=3.3in]{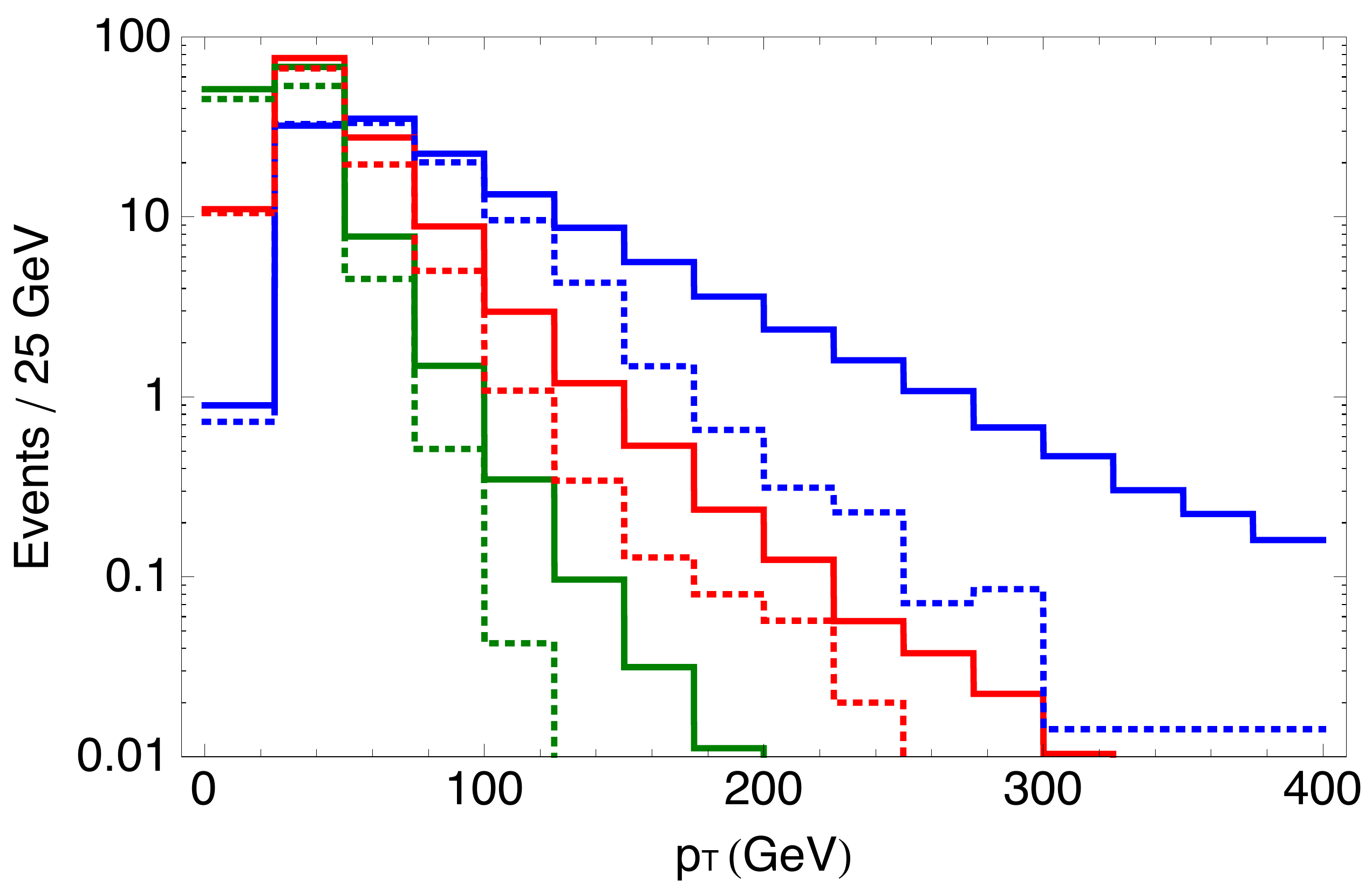}
\caption{$p_T$ spectrum for the 1st, 2nd and 3rd hardest jets in the events (blue, red and green, respectively) when ISR matching is included (solid lines) {\it vs} when all radiation is accounted for by the parton shower (dashed lines). An integrated luminosity of $\mathcal{L} = 70~\text{nb}^{-1}$ is assumed, as well as masses $m_{\tilde{g}}=100\GeV$ and $m_{\chi^0}=90\GeV$.}
\label{Fig: pTs}
\end{center}
\end{figure}

In this work, the efficiencies of the cuts applied by ATLAS' recent search are extracted through a Monte Carlo study.  These efficiencies depend on $m_{\tilde{g}}$ and $m_{\chi^0}$ and are necessary to calculate limits. The signal is calculated using {\tt MadGraph 4.4.32} \cite{Alwall:2007st}, matching parton shower (PS) to additional radiation generated through matrix elements (ME) using the shower $k_{\perp}$ matching prescription from \cite{Alwall:2008qv}, which closely mimics the MLM matching scheme \cite{Alwall:2007fs}. The shower $k_{\perp}$ scheme allows for the matching scale, $Q_{\text{cut}}$, to be set equal to the matrix element cutoff scale, $Q^{\text{ME}}_{\text{cut}}$, and it more directly samples the Sudakov form factor used in the parton shower. In the region where $\tilde{g}$ and $\chi^0$ are nearly-degenerate, the additional radiation is crucial in determining the shape of the $\MET$ distribution and hence how efficiently the signal is found.    A matching scale of  $Q_{\text{cut}} = 100 \GeV$ is adopted for the signal and the matrix elements for the following subprocesses are generated: $2\tilde{g}+0j$, $2\tilde{g}+1j$ and $2\tilde{g}+ 2^+j$.  When performing MLM matching all higher multiplicity jet events are generated through parton showering. The choice of $\alpha_s(\mu)$ in MLM matching is not varied.  Future studies should document the effect of scale choice in matching and incorporate
this theoretical uncertainty into limits.

The importance of matching in describing hard radiation is particularly pronounced for models with nearly degenerate spectra. That is illustrated in Figure \ref{Fig: pTs} for a benchmark mass point $m_{\tilde{g}}=100\GeV$ and $m_{\chi^0}=90\GeV$, which contrasts the $p_T$ spectrum of the 3 hardest jets in the events when matching is included {\it vs} when all radiation is due to the parton shower only.

The parton showering is performed in {\tt PYTHIA 6.4} \cite{Sjostrand:2006za}.  {\tt PYTHIA} also decays $\tilde{g} \rightarrow q\bar{q}\chi^0$, hadronizes the events and produces the final exclusive events.  These events are then clustered using a cone-jet algorithm with $R=0.7$ with {\tt PGS4} \cite{PGS} which also performs elementary fiducial volume cuts and modestly smears the jet energy using the ATLAS-detector card. 

ATLAS' search does not use proper $\MET$ in their analysis,  instead it uses missing energy at the electromagnetic scale, $\MET{}_{\text{\;EM}}$.  The relation between $\MET$ and $\MET{}_{\text{\;EM}}$ is shown in  Fig. 7 of \cite{ATLASMET}. In this paper, it is approximated as
\begin{eqnarray}
\label{Eq: METEM}
\MET{}_{\text{\;EM}} \simeq \frac{ \MET}{ 1.5 - H_T/ 2100\GeV},
\end{eqnarray}
where $H_T$ is the sum of the transverse energies of the jets in the event.  This effectively raises the $\MET$ cut to approximately 50 GeV.

\begin{table*}
\begin{tabular}{|c||l||c|c|c|c|}
\hline
Cut&Topology& $1j+\smMET$& $2^+j+\smMET$& $3^+j+\smMET$& $4^+j+\smMET$\\
\hline\hline
1&$p_{T1}$& $> 70\GeV$&$> 70\GeV$&$> 70\GeV$&$> 70\GeV$\\
\hline
2&$p_{Tn}$& $\le 30\GeV$& $>30\GeV (n=2)$& $>30\GeV (n=2,3)$&$>30\GeV (n=2-4)$\\
\hline
3&$\smMET{}_{\;\text{EM}}$ & $>40\GeV$& $>40\GeV$& $>40\GeV$& $>40\GeV$\\
\hline
4& $p_{T\,\ell}$& $\le 10 \GeV$ & $\le 10 \GeV$ & $\le 10 \GeV$ & $\le 10 \GeV$ \\
\hline
5&$\Delta\phi(j_n, \smMET{}_{\;\text{EM}})$& none& $[>0.2, >0.2]$& $[>0.2, >0.2,>0.2]$&$[>0.2, >0.2,>0.2,\text{none}]$\\
\hline
6&$\smMET{}_{\;\text{EM}}/M_{\text{eff}}$& none& $>0.3$& $>0.25$& $>0.2$\\
\hline\hline
&$N_{\text{Pred}}$& $46^{+22}_{-14}$& $6.6\pm 3.0$& $1.9\pm 0.9$& $1.0\pm0.6$\\
\hline
&$N_{\text{Obs}}$ & 73& 4& 0&1\\
\hline
&$ \epsilon\times\sigma(pp\rightarrow \tilde{g}\tilde{g}X)|_{95\% \text{ C.L.}}$ \;& 663 pb & 46.4 pb & 20.0 pb& 56.9 pb\\
\hline
\end{tabular}
\caption{Searches in \cite{ATLASSusy} used to set limits in this article.  The 95\% C.L. on the production cross section times efficiency of the cuts, $ \epsilon\times\sigma(pp\rightarrow \tilde{g}\tilde{g}X)$, follow from folding in the uncertainties in the luminosity and background.
\label{Tab: Searches}
}
\end{table*}

In order to validate this modeling of $\MET{}_{\text{\;EM}}$, the SU4 mSUGRA model shown in \cite{ATLASSusy} is reproduced.  The  SUSY Les Houches Accord parameter card \cite{SLHA} for SU4 is created with a spectrum calculated with {\tt SuSpect 2.41} \cite{SuSpect} which matches the spectrum reported in \cite{ATLASBigBook} to 5\% accuracy.  The decay card for SU4 is calculated with {\tt SDECAY} \cite{SDECAY}, interfaced with {\tt SUSY-HIT} \cite{SUSYHIT}, and finally the cross sections are generated with {\tt MadGraph} and decayed, showered and hadronized in {\tt PYTHIA}.    The total SUSY production cross section is normalized to the NLO value used in \cite{ATLASSusy} in order to compare efficiencies and shapes of distributions.    

The $\MET{}_{\text{\;EM}}$ distribution for the SU4 benchmark model calculated with {\tt PGS4} is  compared against the $\MET{}_{\text{\;EM}}$ distribution in \cite{ATLASSusy}.  Using the $\MET{}_{\text{\;EM}}$ parametrization in Eq. \ref{Eq: METEM}, the normalization of the {\tt PGS4} distribution is lower by 7\% than the corresponding values in Fig. 10a of \cite{ATLASSusy}. The discrepancy in the individual bins of the $\MET{}_{\text{\;EM}}$ distributions can be quantified by the RMS variance,
\begin{equation}
\Delta=\sqrt{\frac{1}{N_{\text{bins}}}\sum\limits_{i=1}^{N_\text{bins}} \left(\frac{n^\text{ATLAS}_i - n^\text{PGS4}_i}{n^\text{ATLAS}_i}\right)^2}\approx0.1,
\end{equation}
where $N_\text{bins}$ is the number of bins, and $n^\text{ATLAS}_i $ ($n^\text{PGS4}_i$) are the number of events in the $i^\text{th}$ bin of the ATLAS (PGS4) distribution. The absolute efficiency of the fixed $\MET{}_{\text{\;EM}}$ cut  (Cut 3 of Table \ref{Tab: Searches}) in \texttt{PGS4} differs from the ATLAS' absolute efficiency by $\pm 12\%$. 

Fig. 8 of \cite{ATLASSusy} shows the fractional $\MET$ distribution for SU4 (used in Cut 6 Table \ref{Tab: Searches}). The fractional $\MET$ is defined as $\MET/M_{\text{eff}}$, where $M_{\text{eff}}$ is given by
\begin{eqnarray}
M_{\text{eff}} =    \MET{}_{\text{\;EM}}+  \sum_{i=1}^n p_{T\, j_i}.
\end{eqnarray}
Here $n$ is the multiplicity of the channel, {\it e.g. } $n=3$ for the  $3^+j+\MET$ channel.
The difference in the absolute normalization of these distributions is 7\% between ATLAS and {\tt PGS4}.
The $\MET{}_{\text{\;EM}}$ parameterization in Eq. \ref{Eq: METEM} tends to underestimate the SU4 fractional $\MET$ distribution, but  reproduces the efficiency of the  fractional $\MET$ cut to 15\% accuracy.   The fractional $\MET$ cut is more sensitive to the modeling of  $\MET{}_{\text{\;EM}}$ and depends upon the exact kinematics of the signal. 
Therefore, ultimately the results of this paper should be validated by the ATLAS collaboration.

The same procedure used for the calibration process is applied to the study of the models with new light colored particles considered here. After the simulated signal events are processed with {\tt PGS4}, the cut efficiencies for the $pp\rightarrow \tilde{g}\tilde{g}X$ processes are inferred. When combined with the results from the ATLAS searches, a limit can be set on the maximum production cross section for a given $m_{\tilde{g}}$ and $m_\chi^0$.

The cuts applied by ATLAS'  early supersymmetry searches \cite{ATLASSusy} are listed in Table~\ref{Tab: Searches}.   There are several uncertainties folded into the limits.   The  first is the measured luminosity, $\mathcal{L}=70\pm 8~\text{nb}^{-1}$.    For the analysis in this paper it is assumed that  the luminosity is described by a Gaussian distribution. The second uncertainty is the background expectation.  Ref.~\cite{ATLASSusy} used background estimates normalized at low $\MET$ to calculate the QCD contribution to the high $\MET$ signal regions.    Due to the large estimated errors, the number of expected background events is assumed to follow  a log-normal distribution.  There is a weak correlation between the background and luminosity errors but for simplicity these uncertainties are taken to be independent.  Both of these uncertainties are convolved  with a Poisson distribution for signal plus background to derive the 95\% confidence limits (C.L.)  on the cross section times efficiency and are also listed in Table \ref{Tab: Searches}.

Since the signal events typically contain four jets, higher multiplicity searches are expected to provide the best limits. But in practice, the $3^+j+\MET$ had the best exclusion power, and the $4^+j+\MET$ channel was the least sensitive among the multijet searches by factors of several because the $p_{T\,j_4}> 30\GeV$ cut is not efficient.  For instance, for a spectrum with  $m_{\tilde{g}}\simeq 300 \GeV$ and $m_{\chi^0}\simeq 0$, the typical energy of each of the four jets is $\sim$100 GeV. However, since the jets are approximately isotropically distributed, the chance of getting all four jets with $p_T$ above 30\% of their momentum is only half as likely as having only three out of the four jets passing this cut.  Moreover, no events were observed in the $3^+j+\MET$ channel versus one event in the $4^+j+\MET$ channel.

Near the degenerate limit, the $2^+j+\MET$ search was the most effective and set the tightest limit.    The $1j+\MET$ search did not provide competitive limits because of the large number of observed events.

\begin{figure}[htb]
\begin{center}
 \includegraphics[width=3.in]{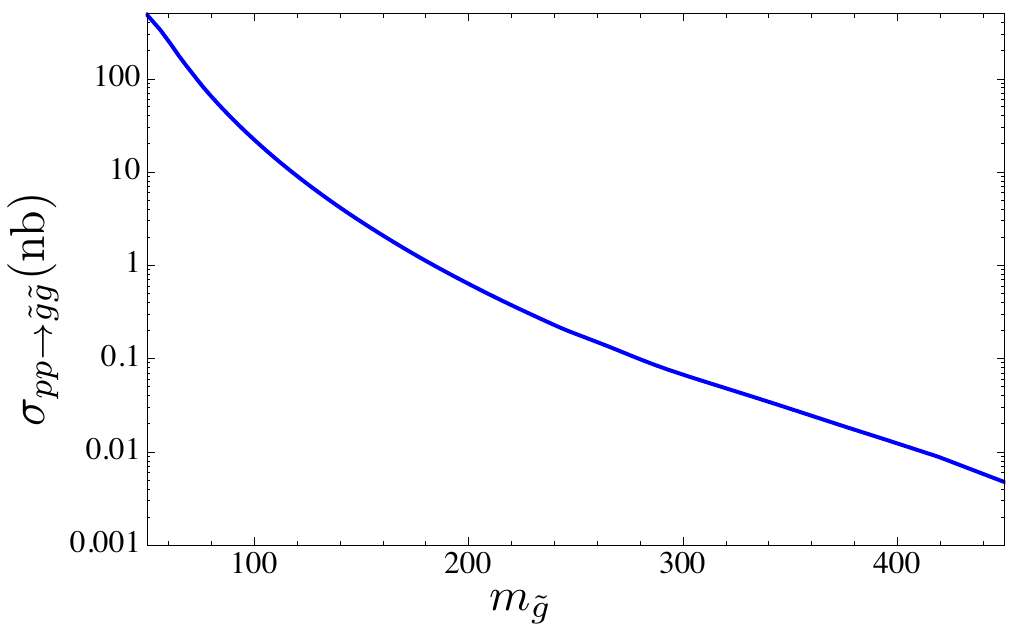}
 \caption{The NLO QCD cross section for $pp\rightarrow \tilde{g}\tilde{g}X$ as a function of the gluino mass.  A cross section of $\sigma = 1/(70 \text{ nb}^{-1})$ corresponds to $m_{\tilde{g}}=395 \GeV$ and $\sigma = 10/(70 \text{ nb}^{-1})$ corresponds to $m_{\tilde{g}}=265 \GeV$.}
   \label{Fig: Cross Section}
   \end{center}
 \end{figure}

The limits placed on the production cross section allow one to infer the excluded parameter space of specific models.  The most straight-forward of those is  the MSSM limit  where the gluino and the LSP are the only currently kinematically accessible particles.  Fig.~\ref{Fig: Cross Section} shows the next-to-leading order (NLO) cross section for $pp\rightarrow \tilde{g}\tilde{g}X$ calculated with \texttt{Prospino 2.0} \cite{Beenakker:1996ed} as a function of the gluino mass.   

\begin{figure}[t]
\begin{center}
 \includegraphics[width=3.4in]{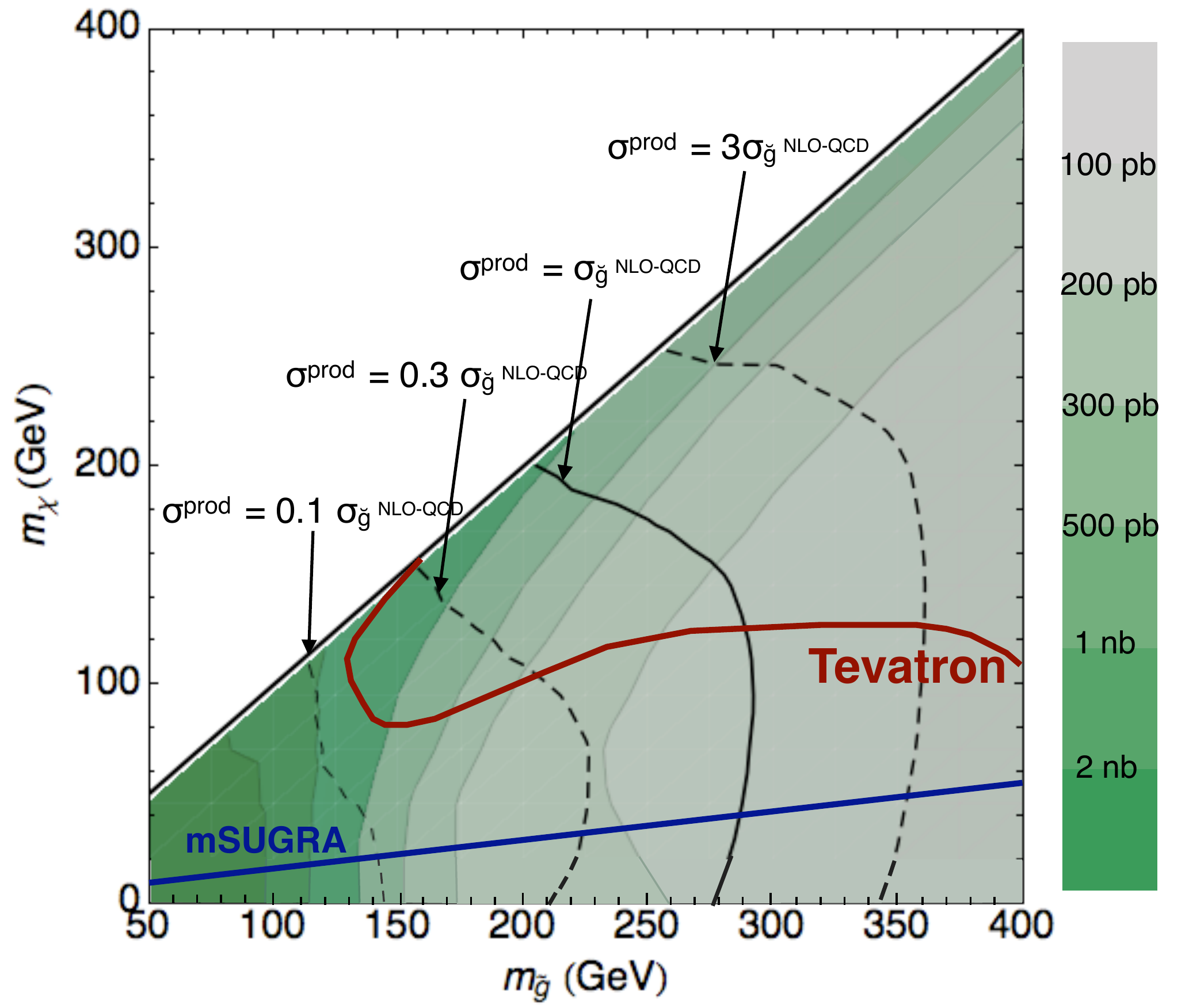}
 \caption{95\% C.L contours of the maximum allowed production cross section $\sigma(pp\rightarrow \tilde{g}\tilde{g}X)$ in the $m_{\tilde{g}}- m_{\chi^0}$ mass plane, for $\tilde{g}$ directly decaying to $\chi^0 j j$.  The contour values are specified in the right color scale.  The dark line corresponds to the exclusion boundary
for models where  the gluino is produced through QCD alone with an NLO cross section (i.e., with all squarks decoupled so that there are no t-channel squark exchange diagrams). The dashed-lines delimit the excluded parameter space of different models where $\sigma(pp\rightarrow \tilde{g}\tilde{g}X)$ is given by a simple rescaling of the NLO-QCD cross section. The red line
  is the current estimate of Tevatron limits  taken from \cite{TevatronJetsMET}.   The blue line denotes a sample mSUGRA spectrum where $\tilde{g}$ is the gluino and $\chi^0$ is the bino. }
   \label{Fig: mGmLSPplot}
   \end{center}
 \end{figure}

The combined results from all the channels are shown in Fig.~\ref{Fig: mGmLSPplot} where only the most restrictive limit on the the production cross section is used.  Fig.~\ref{Fig: mGmLSPplot} displays the contours of the maximum allowed production cross section in the $m_{\tilde{g}}$-$m_{\chi^0}$ parameter space.
 There is a slight loss of sensitivity as $m_{\chi^0}\rightarrow 0$ while holding $m_{\tilde{g}}$ fixed because  the cut on the fractional missing energy has lower efficiencies for larger mass splittings.    This is because as  $\chi^0$ becomes lighter, $\MET$ grows, but the sum of the jet energies grows faster, thereby decreasing the fractional missing energy.

\begin{figure}[htb]
\begin{center}
 \includegraphics[width=3.4in]{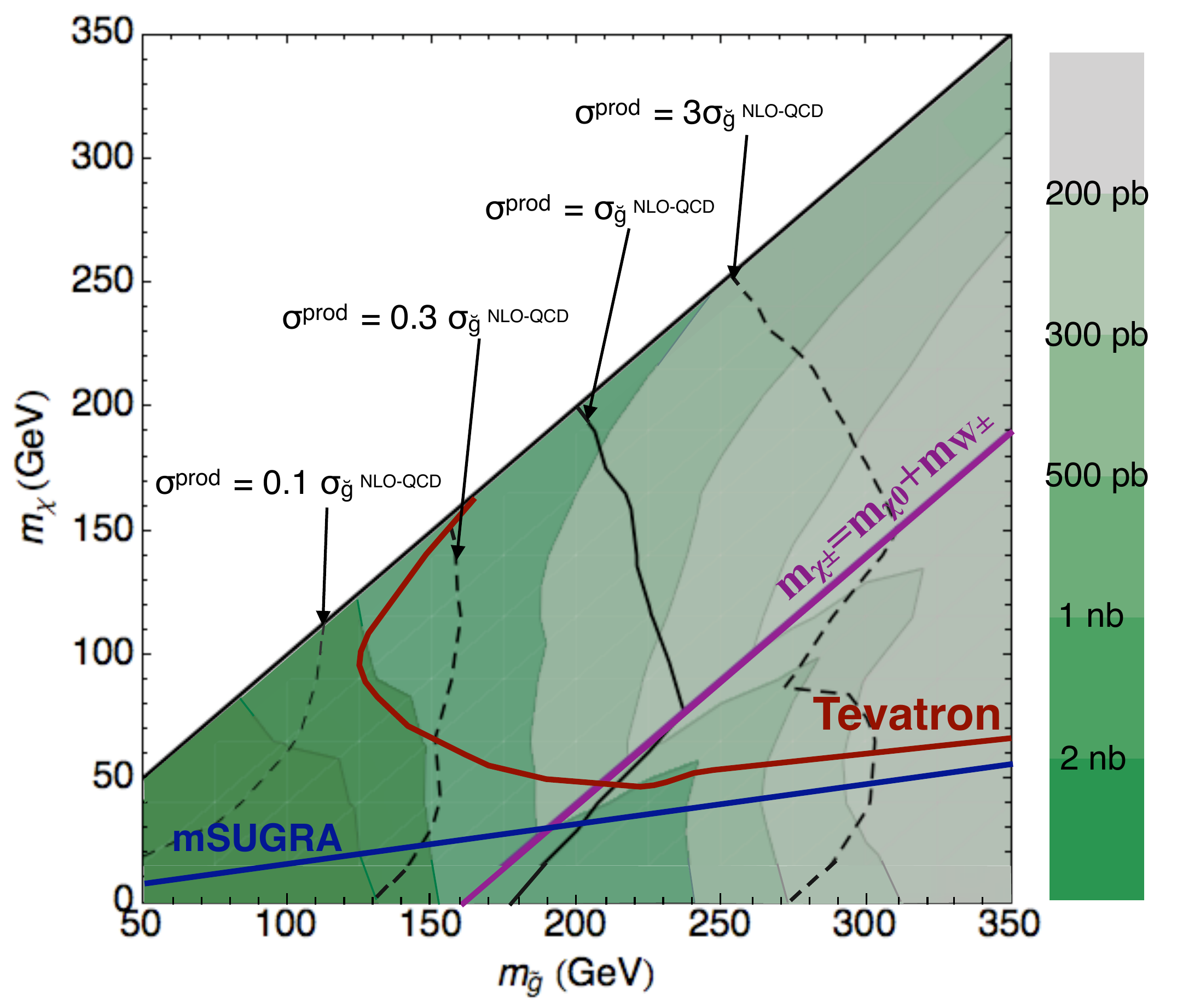}
 \caption{95\% C.L. contours of the maximum allowed production cross section $\sigma(pp\rightarrow \tilde{g}\tilde{g}X)$ in the   $m_{\tilde{g}}- m_{\chi^0}$ mass plane, for a one-step cascade decaying gluino, $\tilde{g}\rightarrow q' \bar{q} \chi^\pm\rightarrow q' \bar{q}  \chi^0 W^{\pm(*)}$.   The contour values are specified in the right color scale.  The purple line corresponds to $m_{\chi^\pm}= m_{\chi^0} + m_{W^\pm}$; below it the $W^\pm$ goes on-shell. 
  The dark line corresponds to the exclusion boundary for models where  the gluino is produced through QCD alone with an NLO cross section (i.e., with all squarks decoupled so that there are no t-channel squark exchange diagrams). The dashed-lines delimit the excluded parameter space of different models where $\sigma(pp\rightarrow \tilde{g}\tilde{g}X)$ is given by a simple rescaling of the NLO-QCD cross section. The red line
  is the current estimate of Tevatron limits taken from \cite{TevatronJetsMET}.   The blue line denotes a sample mSUGRA spectrum where $\tilde{g}$ is the gluino and $\chi^0$ is the bino.   }
   \label{Fig: CascadePlot}
   \end{center}
 \end{figure}

The searches performed by ATLAS also allow new bounds to be placed on pair produced color octets that decay through a one step cascade into jets plus $\MET$. However, the limits have reduced sensitivity because cascade decays reduce the $\MET$ in the events \cite{Cascades}.    
Consider the following simplified model as a way of studying the effects of cascade decays on the search for color octets.
 As before, there is a color octet Majorana fermion, $\tilde{g}$.  In addition to a neutral, Majorana fermion $\chi^0$, there is an intermediate charged fermion, $\chi^\pm$.   The $\chi^\pm$ decays to $\chi^0 W^{\pm(*)}$, where the $W^\pm$ is either on-shell or off-shell  depending on whether the mass splitting between $\chi^\pm$ and $\chi^0$ is greater than or less than $m_{W^\pm}$.   $\tilde{g}$ decays with 100\% branching ratio to $q' \bar{q} \chi^\pm$, with equal probabilities for $\chi^+$ and $\chi^-$.  
 The mass of $\chi^\pm$ is taken to be the arithmetic average of $m_{\tilde{g}}$ and $m_{\chi^0}$, $m_{\chi^\pm} = (m_{\tilde{g}}+ m_{\chi^0})/2$.  This choice of spectrum is  conservative because it reduces the $\MET$ nearly maximally.
 Fig.~\ref{Fig: CascadePlot} shows the limits on this simplified model.  When the $W^\pm$ goes on-shell, the energy put into invisible states decreases and there is a precipitous loss of sensitivity.  The ATLAS search improves the limits in this case as well, placing a universal bound of $m_{\tilde{g}} \simeq 205\GeV$ on the gluino mass.

 \section{Discussion}
 \label{Sec: Outlook}

Based on ATLAS' recent supersymmetry search, new limits for a color octet particle that decays into jets and a missing neutral particle were calculated.  Surprisingly, with remarkably low luminosity these searches surpass the previous limits from Tevatron's analyses. This is possible due to a combination of  the low cuts on missing energy, the quality of the detectors and the level of understanding that the experiments have already gained on jet energy calibration.  
In the future, $b$-jets plus $\MET$ will be another important subclass of topologies to search for and may quickly improve limits over the Tevatron.

This study of the $70~\text{nb}^{-1}$ ATLAS' search applies to many supersymmetric models with a light gluino in the spectrum (see \cite{SWOP} for a more general discussion of limits on non-mSUGRA-like supersymmetric theories).
It also applies to many non-supersymmetric models with a dark matter particle such as  theories of five dimensional  UEDs,  six-dimensional UEDs with scalar adjoints \cite{6DUED}, and little Higgs models with additional colored composites \cite{LH}.  
    
 While color octets that decay into jets plus missing energy are a common feature in models of new physics, there are numerous nearby models that also merit study.
 These include color triplets that decay into a single jet and missing energy, or new colored particles that cascade decay down to the dark matter.  
 The results presented here are applicable to these alternative scenarios as well; in fact, they should apply to any theory where pair-produced new particles each decay to two jets plus a long lived neutral particle, so long as the spectrum is not approximately degenerate. In the degenerate limit, additional QCD radiation in the events becomes important and therefore the color of the new particle is relevant.

The LHC is a huge leap forward and even early analyses  can provide coverage of regions that have not been covered by the Tevatron or earlier experiments.  Ultimately the exact results in this article need to be reproduced and refined by ATLAS to account for detector effects.    CMS also has great promise in this channel and their particle flow algorithms give spectacular measurements of jet energy.  For instance, even CMS' calibration data \cite{CMSDijetsMET} is able to slightly extend the Tevatron's limits.  Going forward,  each experiment has accumulated roughly $1\text{ pb}^{-1}$ of integrated luminosity and therefore there are great possibilities for discoveries in 2010 as the LHC's reach rapidly expands into new kinematic regimes.  

\section*{Acknowledgements}

We are grateful to Andy Haas for useful discussions on ATLAS, the subtleties of missing energy, and statistics; and to James Jackson for useful discussions on CMS.
We also would like to thank  Mariangela Lisanti, Philip Schuster, and Natalia Toro for helpful feedback and Jared Kaplan for reading a preliminary version of the draft.  DSMA, EI and JGW are supported by the DOE under contract DE-AC03-76SF00515. JGW is partially supported by the DOE's Outstanding Junior Investigator Award.


\begin{thebibliography}{99}

\bibitem{ATLASSusy}
ATLAS Collaboration, {\bf ATLAS CONF-2010-065} (2010).

\bibitem{Kaplan:2008pt}
  D.~E.~Kaplan and M.~D.~Schwartz,
 [arXiv:0804.2477 [hep-ph]].


\bibitem{CMSSM}
  S.~Dimopoulos and H.~Georgi,  Nucl.\ Phys.\  B {\bf 193}, 150 (1981).
  R.~Barbieri, S.~Ferrara, C.~A.~Savoy,
  Phys.\ Lett.\  {\bf B119}, 343 (1982).
  N.~Ohta,
  Prog.\ Theor.\ Phys.\  {\bf 70}, 542 (1983).
  L.~J.~Hall, J.~D.~Lykken and S.~Weinberg,
  Phys.\ Rev.\  D {\bf 27}, 2359 (1983).
H. P. Nilles, Phys. Rep. \textbf{110}, 1(1984).
  S.~P.~Martin,
  arXiv:hep-ph/9709356.

\bibitem{Tevatron}
  S.~M.~Wang  [CDF and D0 Collaborations],
  AIP Conf.\ Proc.\  {\bf 1078}, 259 (2009).
  A.~Abulencia {\it et al.}  [CDF Collaboration],
  Phys.\ Rev.\ Lett.\  {\bf 97}, 171802 (2006)
  [arXiv:hep-ex/0605101].
  V.~M.~Abazov {\it et al.}  [D0 Collaboration],
  Phys.\ Lett.\  B {\bf 660}, 449 (2008)
  [arXiv:0712.3805 [hep-ex]].

\bibitem{UA2}
  J.~Alitti {\it et al.}  [UA2 Collaboration],
  Phys.\ Lett.\  B {\bf 235}, 363 (1990).

\bibitem{MirageMediation}
  K.~Choi, K.~S.~Jeong and K.~i.~Okumura,
  JHEP {\bf 0509}, 039 (2005)
  [arXiv:hep-ph/0504037].
  W.~S.~Cho, Y.~G.~Kim, K.~Y.~Lee, C.~B.~Park and Y.~Shimizu,
  JHEP {\bf 0704}, 054 (2007)
  [arXiv:hep-ph/0703163].
  A.~Falkowski, O.~Lebedev and Y.~Mambrini,
  JHEP {\bf 0511}, 034 (2005)
  [arXiv:hep-ph/0507110].
  H.~Baer, E.~K.~Park, X.~Tata and T.~T.~Wang,
  JHEP {\bf 0608}, 041 (2006)
  [arXiv:hep-ph/0604253].
  
  \bibitem{UED}
  T.~Appelquist, H.~C.~Cheng and B.~A.~Dobrescu,
  Phys.\ Rev.\  D {\bf 64}, 035002 (2001)
  [arXiv:hep-ph/0012100].
  H.~C.~Cheng, K.~T.~Matchev and M.~Schmaltz,
  Phys.\ Rev.\  D {\bf 66}, 036005 (2002)
  [arXiv:hep-ph/0204342].

  
\bibitem{TevatronJetsMET}
  J.~Alwall, M.~P.~Le, M.~Lisanti and J.~G.~Wacker,
  Phys.\ Lett.\  B {\bf 666}, 34 (2008)
  [arXiv:0803.0019 [hep-ph]].
  J.~Alwall, M.~P.~Le, M.~Lisanti and J.~G.~Wacker,
  Phys.\ Rev.\  D {\bf 79}, 015005 (2009)
  [arXiv:0809.3264 [hep-ph]].
   
    
\bibitem{Alwall:2007st}
  J.~Alwall {\it et al.},
  JHEP {\bf 0709}, 028 (2007)
  [arXiv:0706.2334 [hep-ph]].

\bibitem{Alwall:2008qv}
  J.~Alwall, S.~de Visscher and F.~Maltoni,
  JHEP {\bf 0902}, 017 (2009)
  [arXiv:0810.5350 [hep-ph]].

\bibitem{Alwall:2007fs}
  J.~Alwall, S.~Hoche, F.~Krauss, N.~Lavesson, L.~Lonnblad, F.~Maltoni, M.~L.~Mangano, M.~Moretti {\it et al.},
  Eur.\ Phys.\ J.\  {\bf C53}, 473-500 (2008).
  [arXiv:0706.2569 [hep-ph]].

\bibitem{Sjostrand:2006za}
  T.~Sjostrand, S.~Mrenna and P.~Z.~Skands,
  JHEP {\bf 0605}, 026 (2006)
  [arXiv:hep-ph/0603175].
  
\bibitem{PGS}
J. Conway, \emph{PGS: Pretty Good Simulator}, http://www.physics.ucdavis.edu/~conway/
research/software/pgs/pgs4-general.htm.


\bibitem{ATLASMET}
ATLAS Collaboration, {\bf ATLAS CONF-2010-057} (2010).

\bibitem{SLHA}
  P.~Z.~Skands {\it et al.},
  JHEP {\bf 0407}, 036 (2004)
  [arXiv:hep-ph/0311123].

\bibitem{SuSpect}
  A.~Djouadi, J.~-L.~Kneur, G.~Moultaka,
  Comput.\ Phys.\ Commun.\  {\bf 176}, 426-455 (2007).
  [hep-ph/0211331].
  
\bibitem{ATLASBigBook}
  G.~Aad {\it et al.}  [The ATLAS Collaboration],
  arXiv:0901.0512 [hep-ex].
  
\bibitem{SDECAY}
  M.~Muhlleitner,
  Acta Phys.\ Polon.\  {\bf B35}, 2753-2766 (2004).
  [hep-ph/0409200].
\bibitem{SUSYHIT}
  A.~Djouadi, M.~M.~Muhlleitner, M.~Spira,
  Acta Phys.\ Polon.\  {\bf B38}, 635-644 (2007).
  [hep-ph/0609292].
  

\bibitem{Beenakker:1996ed}
  W.~Beenakker, R.~Hopker, M.~Spira,
   [hep-ph/9611232].
  
  
  
\bibitem{CDFMonojet}
  T.~Aaltonen {\it et al.}  [CDF Collaboration],
  Phys.\ Rev.\ Lett.\  {\bf 101}, 181602 (2008)
  [arXiv:0807.3132 [hep-ex]].
  A.~Abulencia {\it et al.}  [CDF Collaboration],
  Phys.\ Rev.\ Lett.\  {\bf 97}, 171802 (2006)
  [arXiv:hep-ex/0605101].



  
\bibitem{Cascades}
  H.~Baer, X.~Tata and J.~Woodside,
  Phys.\ Rev.\ Lett.\  {\bf 63}, 352 (1989).
  E.~Izaguirre, M.~Manhart, J.~G.~Wacker,
  [arXiv:1003.3886 [hep-ph]].


\bibitem{SWOP}
  C.~F.~Berger, J.~S.~Gainer, J.~L.~Hewett and T.~G.~Rizzo,
  JHEP {\bf 0902}, 023 (2009)
  [arXiv:0812.0980 [hep-ph]].


  
\bibitem{6DUED}
  B.~A.~Dobrescu, K.~Kong, R.~Mahbubani,
  Phys.\ Lett.\  {\bf B670}, 119-123 (2008).
  [arXiv:0709.2378 [hep-ph]].
  
\bibitem{LH}
  E.~Katz, J.~y.~Lee, A.~E.~Nelson and D.~G.~E.~Walker,
  JHEP {\bf 0510}, 088 (2005)
  [arXiv:hep-ph/0312287].
  T.~Gregoire and E.~Katz,
  JHEP {\bf 0812}, 084 (2008)
  [arXiv:0801.4799 [hep-ph]].


 \bibitem{CMSDijetsMET}
CMS Collaboration, {\it CMS PAS} {\bf JME-10-004} (2010).
CMS Collaboration, {\it CMS PAS} {\bf SUS-10-001} (2010).


 
\end{thebibliography}
\end{document}